\documentclass[12pt]{article}
\usepackage{amsfonts}
\usepackage{bm}
\usepackage{graphicx}
\usepackage{multicol}

\begin{document}

\title{Time correlation functions and Fisher zeros for q-deformed Bose gas
}
\maketitle

\centerline {Kh. P. Gnatenko \footnote{E-Mail address: khrystyna.gnatenko@gmail.com}, A. Kargol \footnote{E-Mail address: akargol@hektor.umcs.lublin.pl}, V. M. Tkachuk \footnote{E-Mail address: voltkachuk@gmail.com}}
\medskip
\centerline {\small  $^{1,3}$ \it  Ivan Franko National University of Lviv, }
\centerline {\small \it Department for Theoretical Physics,12 Drahomanov St., Lviv, 79005, Ukraine}

 \centerline {  \small  $^{2}$ \it Instytut Matematyki, Uniwersytet Marii Curie-Sklodowskiej,}
\centerline {\small \it 20-031 Lublin, Poland}

\abstract{The time-dependent correlation functions of  q-deformed Bose gas are studied.
We find relation of zeros of the correlation functions with the Fisher zeros of partition function of the system.
Complex temperature appears as a result of q-deformation and evolution of correlation function.
A particular case of q-deformed Bose particles on two levels is examined and
 zeros of correlation functions and Fisher zeros of partition function are analyzed.
}

\maketitle

\section{Introduction}

Studies of partition function zeros are important fundamentally. Due to papers \cite{Yang52,Lee52,Fish65} the studies become widely used in examinations of thermodynamical properties of many-body systems, in considerations of phase transitions in various physical systems \cite{Wu}. Also, zeros of partition function fully determine the analytic properties of free energy of a system.

In the contrast to the case of hamiltonian of a system with real parameters, in the case when the parameters in hamiltonian are allowed to be complex the partition function may has zeros which after works of Lee and Yang \cite{Yang52,Lee52} are called Lee-Yang zeros.

 In paper \cite{Lee52} Lee and Yang considered ferromagnetic Ising model with complex magnetic field and found zeros of partition function of the system. The authors proved the theorem that all zeros are purely imaginary. In \cite{Lieb81}
Lieb and Sokal  proved that the Lee-Yang theorem
holds for any Ising-like model with ferromagnetic interaction  (see
also \cite{Koz97,Koz03,Koz99}). The results of Lee and Yang were generalized to the case of complex temperature by Fisher in his paper \cite{Fish65}.

There are difficulties with realization of many-body system with complex parameters at experiment. Therefore for a long time the Lee-Yang zeros were only theoretically studied. The experimental access to the statistical theory of Lee and Yang was provided in \cite{Binek98} and the density function of zeros on the Lee-Yang circle was determined for
a ferromagnet. Then five years ago it was shown that it is possible to observe the Lee-Yang zeros of partition function of spin system in experiment \cite{Wei12,Wei14}.  In  paper \cite{Peng15} the authors reported the experimental observation of Lee-Yang zeros.

 Within the framework of studies of the Lee-Yang zeros the analysis of the dynamical phase transitions was done
 \cite{Flindt13}. In recent paper \cite{Bran17} the experimental determination of the dynamical Lee-Yang zeros was reported.

We would like to mention that there are many papers devoted to studies of
 zeros of partition function for spin systems (see, for instance, \cite{Wei12,Peng15,Kra15,Kra16} and references therein). At the same time zeros of partition function of Bose system and Fermi system are not widely studied (see, for instance, \cite{Mul01,Dij15,Borrmann,Bha11,Zvyagin}).

In our recent
paper \cite{Gna17} we found relation of Lee-Yang zeros of partition function of Bose system with experimentally
observable quantities, namely, with zeros of two-time correlation functions. This relation in principle allows experimental observation of zeros of partition function of Bose system.

In present paper we study the correlation functions of q-deformed Bose gas which is based
on q-deformation of the canonical commutation relation that play important role in different branches
of physics and mathematics (see for instance \cite{QuesnePen03,QuesneTka03,Bor08,Gav16} and references therein).

The paper is organized as follows. In Section 2 we give a preliminary information on q-deformed Bose gas.
In Section 3 we find the relation of zeros of correlation function of q-deformed Bose gas with Fisher zeros of
partition function. Section 4 is devoted to analysis of zeros o correlation function and Fisher zeros in the particular case of Bose particles on two levels.  Conclusions are presented in Section 5.

\section{q-deformed Bose gas}

We consider a system of
$N$ q-deformed Bose particles placed on $s$ levels $\epsilon_i$ ($i=1,2,...s$) and described by the following hamiltonian (see, for instance, \cite{Su93,Man93,Chang02})
\begin{eqnarray}\label{H}
H=\sum_{i=1}^s\epsilon_i \hat a^+_i\hat a_i,
\end{eqnarray}
where creation and annihilation operators of boson on the $i$-th level
$\hat a^+_i$, $\hat a_i$ satisfy the $q$-deformed commutation relations
\begin{eqnarray}
[\hat a_i,\hat a^+_i]_q=\hat a_i\hat a^+_i - q \hat a^+_i\hat a_i=1.
\end{eqnarray}
Operators, corresponding to different levels, commute.
Operator of number of particles $\hat n_i$ on the $i$-th level satisfies the following commutation relations
\begin{eqnarray}
[\hat a_i, \hat n_i]=\hat a_i, \ \ [\hat a^+_i, \hat n_i]=-\hat a^+_i.
\end{eqnarray}

The operators $\hat a_i$, $\hat a^+_i$ and $\hat n_i$ can be represented as
\begin{eqnarray}
\hat a_i=\hat b_i f(\hat b^+_i\hat b_i), \ \ \hat a^+_i=f(\hat b^+_i\hat b_i)\hat b^+_i,  \ \ \hat n_i=\hat b^+_i\hat b_i,
\end{eqnarray}
where
\begin{eqnarray}\label{qnumber}
f(x)={\sqrt {[x]_q\over x}}, \ \ [x]_q={q^x-1\over q-1},
\end{eqnarray}
operators $\hat b_i$, $\hat b^+_i$ and occupation number operator $\hat n_i$ satisfy the ordinary commutation relations
\begin{eqnarray}
[\hat b_i,\hat b^+_j]=\delta_{ij}, \ \ [\hat b_i, \hat n_i]=\hat b_i, \ \ [\hat b^+_i, \hat n_i]=-\hat b^+_i.
\end{eqnarray}
One can easy find eigenstates $|n_1,n_2,...,n_s>$ and corresponding energy levels of hamiltonian (\ref{H})
\begin{eqnarray}
E_{n_1,n_2,...,n_s}=\sum_{i=1}^s\epsilon_i[n_i]_q,
\end{eqnarray}
where $n_i$ are occupation numbers, $n_i=0,1,2,...$, and $[n_i]_q$ is q-deformed number which is defined in (\ref{qnumber}).
Note that we consider canonical ensemble with  fixed number of  Bose particles  $N$, therefore occupation numbers $n_i$ satisfy the following condition
\begin{eqnarray}\label{cond}
\sum_{i=1}^s n_i=N.
\end{eqnarray}

In the next section we consider time-dependent correlation functions of the q-deformed Bose gas and find relation of zeros of the functions with Fisher zeros of partition function.

\section{Correlation functions and Fisher zeros}

Let us consider the following correlation functions of  q-deformed Bose system
\begin{eqnarray}\label{aa}
\langle \prod_{j=1}^s\hat a^+_j(t_1) \hat a_j(t_2) \rangle=\nonumber\\=
{1\over Z(\beta)}{\rm Tr} e^{-\beta H} \hat a^+_1(t_1) \hat a_1(t_2)...\hat a^+_s(t_1) \hat a_s(t_2),
\end{eqnarray}
where $Z(\beta)$ is partition function
\begin{eqnarray}
Z(\beta)={\rm Tr} e^{-\beta H},
\end{eqnarray}
here $\beta=1/kT$ is the inverse temperature and
\begin{eqnarray}
\hat a_j(t)=e^{iHt/\hbar}\hat a_je^{-iHt/\hbar}.\label{at23}
\end{eqnarray}
Substituting (\ref{H}) into (\ref{at23}), we obtain

\begin{eqnarray}\label{ajt}
\hat a_j(t)=e^{i\epsilon_j \hat a^+_j\hat a_j t/\hbar} \hat a_j e^{-i\epsilon_j \hat a^+_j\hat a_j t/\hbar }.
\end{eqnarray}

In order to rewrite $\hat a_j(t)$ in the form which is convenient for calculation of the correlation functions,  we use the following identities
\begin{eqnarray}
\hat a_j f(\hat a^+_j \hat a_j)=f(\hat a_j \hat a^+_j)\hat a_j=f(q\hat a^+_j \hat a_j+1)\hat a_j, \\
f(\hat a^+_j \hat a_j)\hat a_j = f(\hat a_j \hat a^+_j/q-1/q)\hat a_j=\nonumber\\=\hat a_j f(\hat a^+_j \hat a_j/q-1/ q),\label{l}
\end{eqnarray}
here $f$ is an arbitrary function for which the Taylor expansion exists.
Using  identity (\ref{l}),
we can rewrite (\ref{ajt}) as follows
\begin{eqnarray} \label{at}
\hat a_j(t)= \hat a_j e^{-i\epsilon_j t/q\hbar}e^{-i(1-1/q)\epsilon_j\hat a^+_j \hat a_j t/\hbar}.
\end{eqnarray}
The conjugated operator to $\hat a_j(t)$ reads
\begin{eqnarray}\label{at+}
\hat a^+_j(t)=e^{i\epsilon_j t/q\hbar}e^{i(1-1/q)\epsilon_j\hat a^+_j \hat a_j t/\hbar}\hat a^+_j.
\end{eqnarray}

Substituting  (\ref{at}) and (\ref{at+}) into (\ref{aa}), we find

\begin{eqnarray}
\langle \prod_{j=1}^s\hat a^+_j(t_1) \hat a_j(t_2) \rangle=\nonumber\\=
{1\over Z(\beta)}{\rm Tr} e^{-\beta H} \prod_{j=1}^s
e^{i\epsilon_j \tau /q\hbar}e^{i(1-1/q)\epsilon_j\hat a^+_j \hat a_j \tau/\hbar}
\hat a^+_j \hat a_j=\nonumber \\
=e^{i\sum_j\epsilon_j \tau /q\hbar}{1\over Z(\beta)}{\rm Tr} e^{-\beta H}
e^{\sum_j i(1-1/q)\epsilon_j\hat a^+_j \hat a_j \tau/\hbar}
\prod_{j=1}^s\hat a^+_j \hat a_j,\nonumber\\
\end{eqnarray}
where $\tau=t_1-t_2$.
Taking into account  (\ref{H}), the expression for correlation function can be rewritten
in the following form
\begin{eqnarray}\label{aatbeta}
\langle \prod_{j=1}^s\hat a^+_j(t_1) \hat a_j(t_2) \rangle=
e^{i\sum_j\epsilon_j \tau /q\hbar}{1\over Z(\beta)}{\rm Tr} e^{-{\tilde \beta} H}
\prod_{j=1}^s\hat a^+_j \hat a_j,
\end{eqnarray}
here we introduce the complex temperature
\begin{eqnarray}\label{cbeta}
{\tilde \beta}=\beta- i(1-1/q) \tau/\hbar=\beta+i\beta_1.
\end{eqnarray}
 We would like to stress that the imaginary part $\beta_1$ of the complex temperature is caused by the q-deformation and is related with the time of evolution.
Note that in the case when deformation is absent, namely $q=1$, the imaginary part of the complex temperature is equal to zero.

This result can be rewritten as
\begin{eqnarray}\label{aaZ}
\langle \prod_{j=1}^s\hat a^+_j(t_1) \hat a_j(t_2) \rangle=\nonumber\\=
e^{i\sum_j\epsilon_j \tau /q\hbar}{1\over Z(\beta)}
\left(-{1\over\tilde\beta}\right)^s \left(\prod_{j=1}^s{\partial\over\partial \epsilon_j}\right)
{\rm Tr} e^{-{\tilde \beta} H}.
\end{eqnarray}
Taking Tr over the eigenstates of hamiltonian (\ref{H}), we find
\begin{eqnarray}\label{Ztb}
Z(\tilde \beta)={\rm Tr} e^{-{\tilde \beta} H}=\nonumber\\=\sum_{n_1=0}^N\sum_{n_2=0}^N\cdots \sum_{n_s=0}^N e^{-\tilde\beta (\epsilon_1[n_1]_q+\epsilon_2[n_2]_q+...+\epsilon_s[n_s]_q)}.
\end{eqnarray}
Here the occupation numbers $n_j$ satisfy condition (\ref{cond}).
Therefore, the sum over the occupation numbers in  (\ref{Ztb}) can not be factorized.

It is important to note that partition function (\ref{Ztb}) contains complex temperature.
Thus we find the relation of correlation function with partition function containing complex temperature (\ref{aaZ}).

\section{q-deformed Bose particles on two levels}

 Let us study a particular case of two-level system of $N$ q-deformed Bose particles which is described by hamiltonian (\ref{H}) with $s=2$. In this case, taking into account condition (\ref{cond}), we have $n_2=N-n_1$. So, partition function (\ref{Ztb}) can be reduced to the following expression
\begin{eqnarray}\label{ZN}
Z(\tilde \beta)=\sum_{n_1=0}^N e^{-\tilde\beta (\epsilon_1[n_1]_q+\epsilon_2[N-n_1]_q)}.
\end{eqnarray}
For correlation function in this case we have
\begin{eqnarray}
\langle \hat a^+_1(t_1) \hat a_1(t_2) \hat a^+_2(t_1) \hat a_2(t_2)\rangle=
e^{i\sum_j\epsilon_j \tau /q\hbar}{1\over Z(\beta)}\times\nonumber\\ \times
\sum_{n_1=0}^N e^{-\tilde\beta (\epsilon_1[n_1]_q+\epsilon_2[N-n_1]_q)}[n_1]_q[N-n_1]_q=\nonumber\\
=e^{i\sum_j\epsilon_j \tau /q\hbar}{Z_c(\tilde\beta)\over Z(\beta)},
\end{eqnarray}
where we introduce the notation
\begin{eqnarray}\label{ZcN}
Z_c(\tilde\beta)=\sum_{n_1=0}^N e^{-\tilde\beta (\epsilon_1[n_1]_q+\epsilon_2[N-n_1]_q)}[n_1]_q[N-n_1]_q.
\end{eqnarray}
Note that zeros of correlation function correspond to zeros of $Z_c(\tilde\beta)$.

Fisher zeros of partition function and zeros of correlation function can be found analitically for
small number of particles.
In the case of $N=1$ the correlation function is zero and the partition function reads
\begin{eqnarray}\label{ZN1}
Z(\tilde\beta)=e^{-\tilde\beta\epsilon_1}\left(1+e^{-\tilde\beta\Delta\epsilon}\right),
\end{eqnarray}
where $\Delta\epsilon=\epsilon_2-\epsilon_1$.
In this case equation for zeros of partition function $Z(\tilde\beta)=0$   has the following solutions
\begin{eqnarray}
\beta=0, \ \ \beta_1\Delta\epsilon=\pi(2n+1), \ \ n=0,\pm 1, \pm 2, ...
\end{eqnarray}
Here $\beta$ and $\beta_1$ are real and imaginary part of $\tilde\beta$.
It is convenient to introduce new variable $z=e^{-\tilde \beta\epsilon_1}$. Then possible zeros lay on the
circle of unit radius in $z$-plane.
Note that according to (\ref{cbeta}) at $q=1$ imaginary part of temperature reads $\beta_1=0$. Therefore in the case when
$q=1$ we have no solutions for zeros of partition function.

In the case of system of two particles one also has  trivial result for correlation function, it is equal to zero.  In the case when $N=3$ zeros of correlation function can be found analitically and they are nontrivial. We have
\begin{eqnarray}\label{ZcN3}
Z_c(\tilde\beta)=(1+q)e^{-\tilde\beta(\epsilon_1(q+1)+\epsilon_2)}\left(1+e^{-\tilde\beta q\Delta\epsilon}\right).
\end{eqnarray}
Zeros of correlation function in this case are achieved at
\begin{eqnarray}\label{CorN3}
\beta=0, \ \ \beta_1 q\Delta\epsilon=\pi(2n+1), \ \ n=0,\pm 1, \pm 2, ...
\end{eqnarray}
From (\ref{cbeta}) and (\ref{CorN3}) we find that correlation function in this case has zeros at times
\begin{eqnarray}\label{tauCor}
\tau={\hbar\over (q-1)\Delta\epsilon}\pi(2n+1), \ \ n=0,\pm 1,\pm2, ...
\end{eqnarray}
The zeros for correlation function in this case lay
on the circle of unit radius in $z$-plane.
For $q=1$ the correlation function does not have finite zeros, namely, zeros of correlation function tend to infinity at $q\to 1$.

We would like to note that in general case of arbitrary number of particles $N$ the zeros of partition function and zeros of correlation function determined on the z-plane ($z=e^{-\tilde\beta\epsilon_1}$) are roots of polynomials with real powers.
\begin{eqnarray}
Z=\sum_{n_1=0}^N z^{[n_1]_q+[N-n_1]_q\epsilon_2/\epsilon_1}, \label{ZZN}\\
Z_c=\sum_{n_1=0}^N z^{[n_1]_q+[N-n_1]_q\epsilon_2/\epsilon_1}[n_1]_q[N-n_1]_q.\label{ZZZN}
\end{eqnarray}

In particular case when the parameter of deformation reads $q=2$
 we have that the q-numbers
\begin{eqnarray}
[n_1]_q=2^{n_1}-1, \ \ [N-n_1]_q=2^{N-n_1}-1
\end{eqnarray}
are integer.  So, when in additional
$\epsilon_2/\epsilon_1$ is integer the
expressions (\ref{ZZN}) and (\ref{ZZZN}) can be rewritten in the form of polynomial over
$z$ with integer powers
\begin{eqnarray}\label{ZNZcN}
Z=\sum_{n_1=0}^N z^{(2^{n_1}-1)+(2^{N-n_1}-1)\epsilon_2/\epsilon_1}, \label{fff1}\\
Z_c=\sum_{n_1=0}^N z^{(2^{n_1}-1)+(2^{N-n_1}-1)\epsilon_2/\epsilon_1}(2^{n_1}-1)(2^{N-n_1}-1).\label{fff2}
\end{eqnarray}
Fisher zeros of partition function and correlation function in this case are presented on Figure 1. Note that the number of Fisher zeros  exceed the number of particles. This is the result of exponential dependence of energy on the number of q-deformed Bose particles on a given level.

\begin{figure}[h!]
\includegraphics[width=0.4\textwidth]{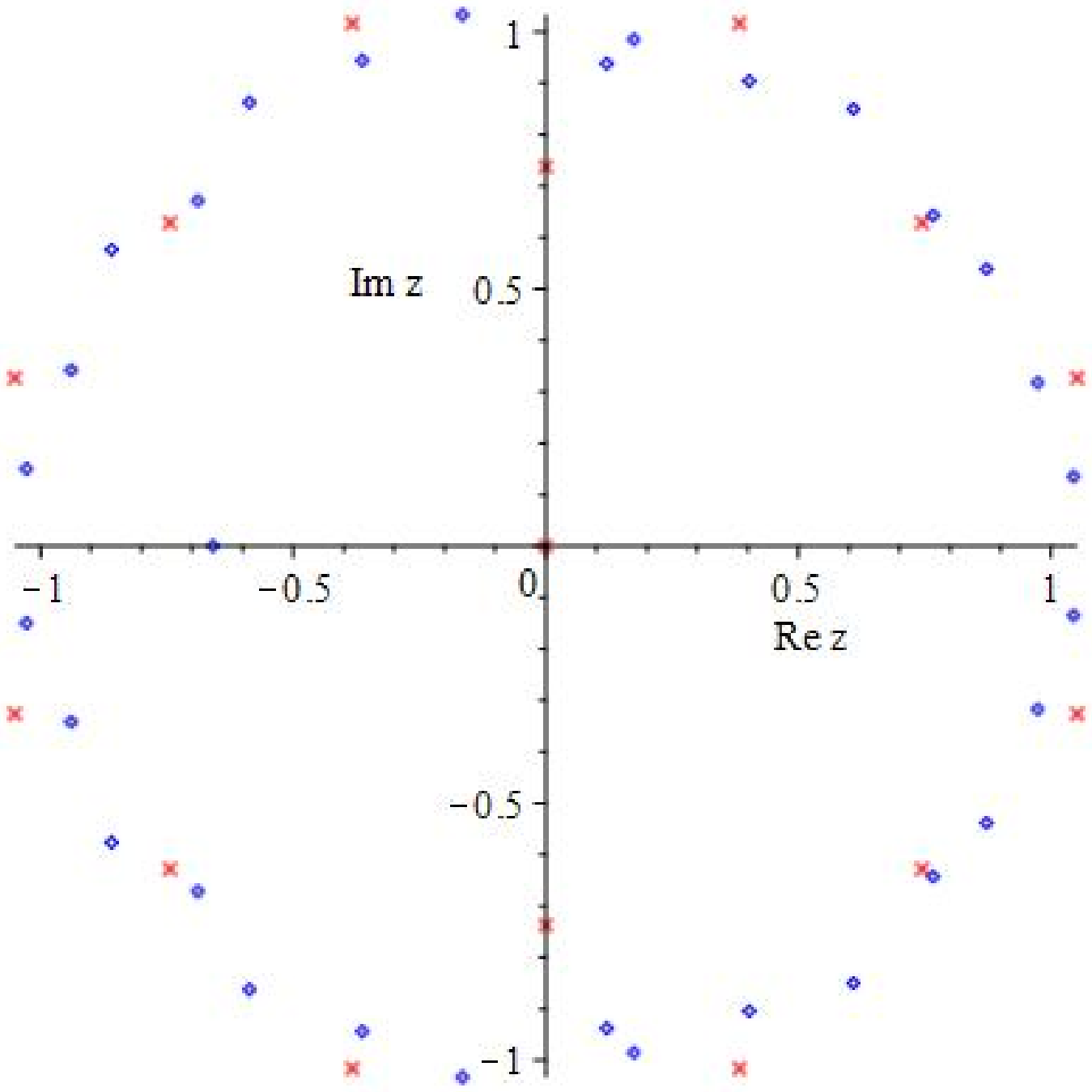}{a}
\quad
\includegraphics[width=0.4\textwidth]{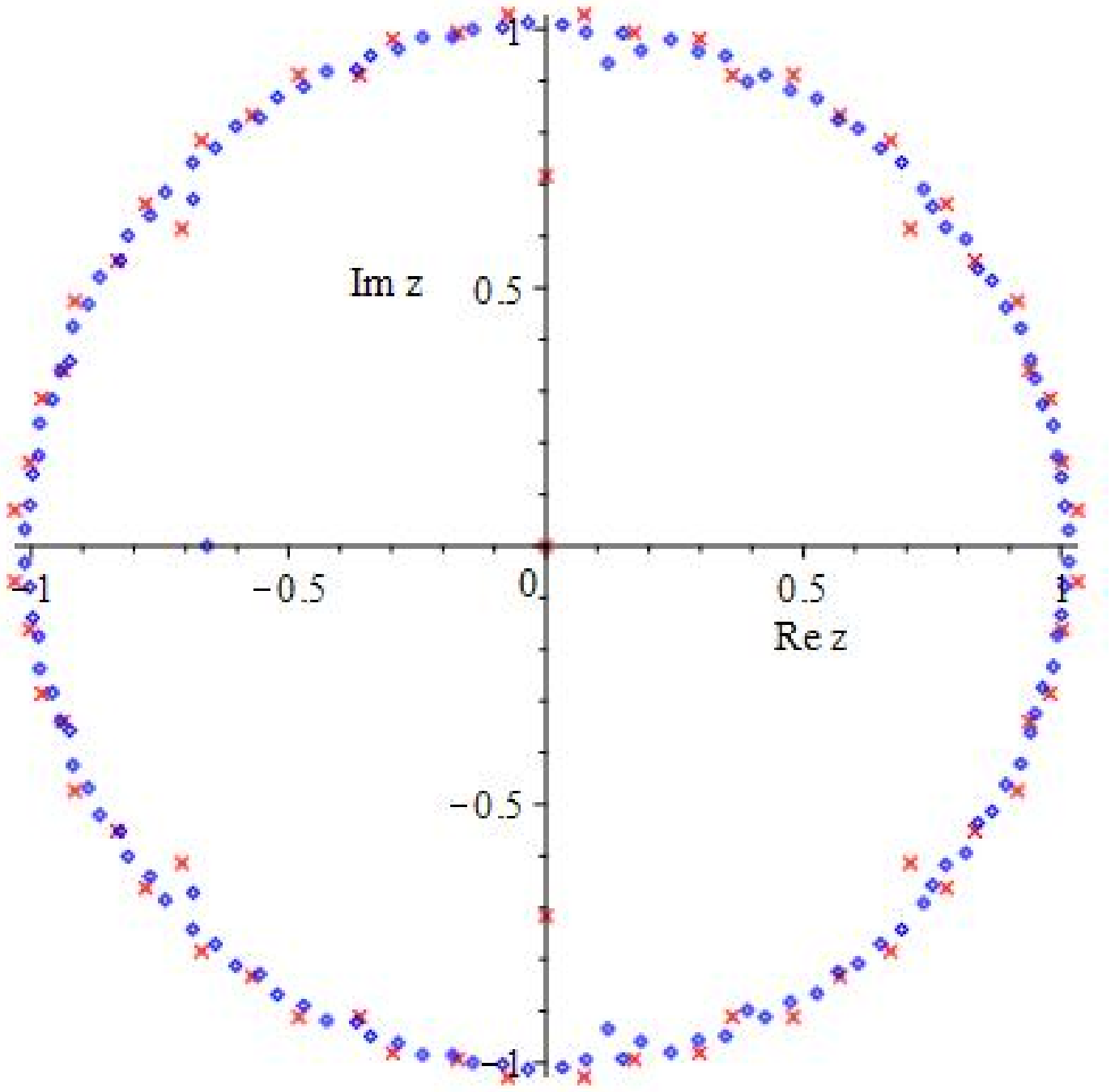}{b}
\caption{ Fisher zeros (\ref{fff1}) (marked by circles)  and  zeros of correlation function (\ref{fff2}) (marked by crosses) for $q=2$, $\epsilon_2=0$ (a) $N=5$, (b) $N=7$.}
\label{f1}
\end{figure}

\section{Conclusion}

In this paper the time-dependent correlation functions of q-deformed Bose gas have been studied.
The main our result is the relation of correlation function of the q-deformed Bose system with partition function which depends on complex temperature.
 Namely, we have found that correlation functions can be represented  as (\ref{aaZ}). So, there is relation of zeros of correlation function of q-deformed Bose gas with the Fisher zeros of partition function. It is important to  note that the complex temperature is caused by q-deformation and evolution of the system. If deformation is absent, namely if $q=1$, the imaginary part of the temperature is equal to zero.

Particular case of system of q-deformed Bose particles on two levels has been examined.
 We have found analytically that zeros of partition function for one particle $N=1$ and zeros of correlation function for $N=3$  lay on the unit circle in $z$-plane which corresponds to purely imaginary zeros in $\beta$-plane. In this case zeros of correlation function are achieved during evolution in the times which are given by (\ref{tauCor}).

 We have also considered the particular case when $q=2$ and the ratio of energies of two levels $\epsilon_2/\epsilon_1$ is integer. In this case the zeros of the partition function and the zeros of correlation function correspond to the roots of polynomial with integer powers. The Fisher zeros of partition function and correlation function for q-deformed Bose gas in the case of $q=2$ for particular numbers of particles of the system are presented on Figure 1. Note that as it is shown on the Figure  the zeros lay close to the unit circle but not exactly on the circle.

 Finally, we would like to note that in contrast to the ordinary Bose gas where interaction is responsible for Lee-Yang zeros (as was shown in \cite{Gna17})  the q-deformation leads to the Fisher zeros.

\section{Acknowledgments}
This work was supported in part by the European Commission under the project STREVCOMS PIRSES-2013-612669, by the State Found for Fundamental Research under the project $\Phi$-76/105-2017 and by the project $\Phi\Phi$-63Hp (No. 0117U007190) from the Ministry of Education and Science of Ukraine. The authors thank Prof. Yu. Kozitsky, Prof. Yu. Holovatch, Dr. M. Krasnytska for useful discussions.

\end{document}